\documentclass[twocolumn,aps,letter]{revtex4}
\usepackage{graphicx,graphics,float}
\usepackage{dcolumn}
\usepackage{bm}
\usepackage{amsmath,amssymb}
\begin{document}
\title{Thermodynamic signatures of an underlying quantum phase transition: A grand canonical approach}
\author{Kevin Jimenez and Jose Reslen}
\affiliation{Coordinaci\'on de F\'{\i}sica, Universidad del Atl\'antico, Kil\'ometro 7 Antigua v\'{\i}a a Puerto Colombia, A.A. 1890, Barranquilla, Colombia.}%
\date{\today}
\begin{abstract}
The grand canonical formalism is employed to study the thermodynamic structure of a model displaying
a quantum phase transition when studied with respect to the canonical formalism. A numerical survey shows that the 
grand partition function diverges following a power law when the interaction parameter approaches
a limiting constant. The power-law exponent takes a distinctive value when such limiting constant coincides
with the critical point of the subjacent quantum phase transition. An approximated expression for the grand 
partition function is derived analytically implementing a mean field scheme and a number of thermodynamic 
observables are obtained. The system observables show signatures that can be used to track
the critical point of the underlying transition. This result provides a simple fact that
can be exploited to verify the existence of a quantum phase transition avoiding the zero temperature regime.
\end{abstract}
\maketitle
\section{Introduction}
\label{intro}
The observation of quantum-interference effects in many-body systems is often
deterred by the very short coherence times displayed by quantum pure-states
in nature. This affects in particular cooperative states resulting from 
interaction-dominated phases in many-body systems. These states have important
applications in quantum computation and nanoelectronics because interaction 
is key to develop control mechanisms. In contrast to pure states, mixed states
are less prone to be demolished by decoherence \cite{Zurek}, especially when they correspond
to equilibrium states because their entropies are maximal and the system
cannot loss any more information to the environment. As a result, it is reasonable
to assume that the observation of specific effects in many-body systems 
through thermodynamic states is feasible, even practical, as long as it be
possible to easily keep the system in equilibrium. 

Quantum Phase Transitions (QPTs) are physical processes arising from a change
in the ground state structure of a system as a parameter crosses a transition-or-critical
point \cite{QFT}. These transitions occur at zero temperature and they are strongly
influenced by quantum correlations. In fact, it is known that the amount of entanglement
present in a system is maximal at, or close to, the critical point of a second order QPT \cite{vedral}.
The universality class of a QPT is determined by the power law exponents 
that define the scaling behavior of characteristic variables in the vicinity of the critical point.
Recently, there has been interest in knowing how correlations, either classical
or quantum, behave at finite temperature in models showing well understood QPTs \cite{Wilms,Brandes}. 
These investigations have been made using canonical ensemble theory: the system is
in thermodynamic equilibrium with a bath at fixed temperature and the number of
particles is fixed. In contrast, applications of the grand-canonical-ensemble 
theory to the same kind of systems are, to the best of the authors' knowledge, not available so far. 
The element that is added in the grand canonical formulation is the notion of fluctuations 
of the number of particles. In this case an open quantum system interacts with a bath in such a way that 
not only energy but also particles can be exchanged. This additional consideration might better describe 
the conditions encountered in some low-temperature and solid-state experiments.
Let us consider a system governed by the following Hamiltonian
\begin{equation}
\hat{H}_M = \hat{a}_1^{\dagger}\hat{a}_2 + \hat{a}_2^{\dagger} \hat{a}_1  -\frac{\lambda}{M} \left ( \hat{a}_1^{\dagger}\hat{a}_1 \hat{a}_1^{\dagger}\hat{a}_1 + \hat{a}_2^{\dagger}\hat{a}_2 \hat{a}_2^{\dagger}\hat{a}_2   \right ).
\label{eq:1}
\end{equation}
The operators that describe the Hamiltonian follow bosonic commuting relations
$[\hat{a}_1,\hat{a}_1^{\dagger}]=[\hat{a}_2,\hat{a}_2^{\dagger}]=1$ 
and $[\hat{a}_1,\hat{a}_2]=0$. Symbols $M$ and $\lambda$ represent the number of particles and 
the intensity of the interaction among bosons respectively. By 
definition $\hat{H}_0=0$. The Hamiltonian has been normalized so 
that $\lambda$ is dimensionless and the energy unit is half the
energy difference between the eigenenergies of $\hat{H}_1$.
In this letter only the case $\lambda>0$ is considered. It is possible to 
change the sign of the single particle
term so that it better resembles a kinetic energy contribution 
applying a unitary transformation producing $\hat{a}_1 \rightarrow i\hat{a}_1$ and
$\hat{a}_2 \rightarrow -i\hat{a}_2$. This scheme can be seen as a simple model describing 
a system of cold atoms tunneling between symmetric adjacent wells and undergoing 
attractive interactions \cite{doublewell,doublemode}. It is known that in actual experiments both
the double-well profile and the interaction intensity can be controlled to
a great degree \cite{Wieman}. Usually, the confining profile is realized using 
counter-propagating laser beams that form a periodic super-lattice while
the interaction can be tuned applying a magnetic field near a Feshbach resonance \cite{Fesh}. 
In numerical studies it is useful to exploit the fact that $\hat{H}_M$ commutes with the following operators,
\begin{equation}
\hat{M} = \hat{a}_1^{\dagger} \hat{a}_1 + \hat{a}_2^{\dagger} \hat{a}_2, \hspace{0.5 cm} \hat{\Pi} = e^{i \frac{\pi}{2} \left ( \hat{a}_1^{\dagger}\hat{a}_2 +  \hat{a}_2^{\dagger} \hat{a}_1  \right )}.
\label{eq:2}
\end{equation}
These commutation properties imply that the eigenstates of (\ref{eq:1}) 
display fixed number of particles and, for non-degenerate spectra, parity. 
This latter symmetry emerges as a consequence of the invariance of the 
Hamiltonian under the swap of labels (wells) 1$\leftrightarrow$2. The system behavior
is determined by the trade-off between hopping and attractive interaction. 
Hamiltonian (\ref{eq:1}) can be written in terms of angular 
momenta through the following Schwinger transformation,
\begin{equation}
\hat{J}_z = \frac{\hat{a}_1^{\dagger}\hat{a}_2 + \hat{a}_2^{\dagger} \hat{a}_1}{2}, \hspace{0.2cm}
\hat{J}_x = \frac{\hat{a}_1^{\dagger}\hat{a}_1 - \hat{a}_2^{\dagger} \hat{a}_2}{2}.
\label{eq:3}
\end{equation}
Inserting these identities in Eq. (\ref{eq:1}) and after a few arrangements we arrive to,
\begin{equation}
 2 \hat{J}_z - \frac{2 \lambda }{M} \hat{J}_x^2 - \frac{M \lambda}{2},
\label{eq:4}
\end{equation}
which corresponds to a particular case of the Lipkin-Meshkov-Glick (LMG) model \cite{Lipkin}.
The Hamiltonian form shown in Eq. (\ref{eq:4}) has been extensively studied with 
reference to, among many others, its scaling behavior \cite{Dusuel,Orus,Colonna}, energy 
spectrum \cite{Ribeiro}, correlations at finite temperature \cite{Wilms} and 
applications to quantum metrology \cite{Salvatori}. If the angular momenta are written 
as sums of spins, $\hat{J}_{x,z} = \frac{1}{2} \sum_{j=1}^M \hat{\sigma}_j^{x,z}$, 
where $\hat{\sigma}^{x,z}$ are Pauli matrices, the model becomes 
\begin{equation}
\sum_{j=1}^M  \hat{\sigma}_j^{z} - \frac{\lambda}{M} \sum_{j=1}^M \sum_{k=1}^{j-1} \hat{\sigma}_k^x \hat{\sigma}_j^x   - M \lambda.
\label{eq:33}
\end{equation}
In this notation, and up to a constant factor, the model is known as the infinite
range Ising model because the interaction among spins is completely homogeneous with
respect to the spin index. It is worth mentioning that Hamiltonian (\ref{eq:1}) is not
completely equivalent to Hamiltonian (\ref{eq:33}), as can be seen by comparing their respective
Hilbert space dimensions. Indeed, the sums of spins give rise to various irreducible 
representations of Hamiltonian (\ref{eq:4}) corresponding to different values of total 
angular momentum. The representation with the biggest total angular momentum corresponds to Hamiltonian (\ref{eq:1}). 
It can be shown that, up to an additive constant proportional to $M$, Hamiltonian (\ref{eq:1}) is the bosonic second-quantization of 
Hamiltonian (\ref{eq:33}) and as such it is spanned by the symmetric states of the spin basis.
This affects the density of states and eventually derives in the fact that Hamiltonians (\ref{eq:4})
and (\ref{eq:33}) exhibit a QPT as well as a phase transition at finite temperature, while 
Hamiltonian (\ref{eq:1}) displays only a QPT. Such a QPT can be studied by 
assuming that the ground state is given as follows \cite{doublewell}
\begin{equation}
|G (\theta) \rangle = \frac{\left. {\hat{b}^{\dagger}}\right.^M |0\rangle}{\sqrt{M!}}, \hspace{0.1cm} \hat{b}^{\dagger} =  \hat{a}_1^{\dagger} \cos \theta-  \hat{a}_2^{\dagger} \sin \theta,
\label{eq:15}
\end{equation}
where $\theta$ is bounded to the interval $[0,\pi]$ in order to avoid redundancies.
The angle $\theta$ takes the value that minimizes the energy
\begin{equation}
E_G = Min_\theta \langle G(\theta) | \hat{H}_M | G(\theta) \rangle.
\label{eq:16}
\end{equation}
After some direct calculations we obtain to leading order in $M$
\begin{equation}
\text{if $\lambda < 1$}, \text{  } \theta^* = \frac{\pi}{4}  \text{ and }  E_G = -M \left( 1 + \frac{\lambda}{2} \right).
\end{equation}
Otherwise
{\small
\begin{equation}
\text{if $\lambda \ge 1$},\text{ } \theta_1^* = \frac{1}{2} \arcsin \left( \frac{1}{\lambda} \right) \text{or} \text{ } \theta_2^* = \frac{\pi}{2} - \theta_1^*,  \end{equation}
}
and $E_G = -M \left( \lambda + \frac{1}{2 \lambda} \right )$. Canonical ensemble theory dictates
that the statistical state becomes $| G(\theta^*) \rangle$ for $\lambda < 1$ and
\begin{equation}
\frac{1}{2} \left( | G(\theta_1^*) \rangle \langle G(\theta_1^*) |  + | G(\theta_2^*) \rangle \langle G(\theta_2^*) |  \right),
\label{eq:gambler}
\end{equation}
for $\lambda \ge 1$. The QPT is characterized by a structural change in the spectrum of the Hamiltonian, which
goes from a gaped phase with non-degenerate energy levels for $\lambda < 1$, to a gapless phase
with a double degeneration of every level \cite{comment1} for $\lambda \ge 1$. Such a change in the density of states
takes place only in the limit $M \rightarrow \infty$ (the thermodynamic limit) and is marked by 
a discontinuity at $\lambda=1$ in the second derivative
of the rescaled free energy \cite{comment2}. As the free energy is continuous at the critical point,
the transition is classified as a second order QPT. Neither $| G(\theta_1^*) \rangle$ nor $| G(\theta_2^*) \rangle$
are invariant under parity transformations, because they display different occupation
numbers at each side of the double well. Contrariwise, both state (\ref{eq:gambler}) 
and $| G(\theta^*) \rangle$ are invariant, and as such it can be said that symmetry is 
preserved across the critical point as long as the system remains in thermodynamic equilibrium
so that the transition be reversible.
The general purpose of this work is to analyze the thermodynamic properties of a system
governed by Hamiltonian (\ref{eq:1}) using the grand canonical formalism, i.e.,
assuming that the number of particles is not fixed but subject to statistical
fluctuations determined by the characteristic conditions of a surrounding bath.
In particular, it is of interest to examine whether signatures of the aforementioned 
QPT can be in any way seen in the resulting framework. The underlying intention is to 
establish a connection of physical significance between the properties of the system 
in the thermodynamic limit and its finite size structure as a whole.
\section{Grand canonical approach}
\label{gca}
The thermodynamics of the model is determined by the grand canonical partition
function,
\begin{equation}
\Xi = \sum_{M=0}^{\infty} Tr \left (  e^{-\beta (\hat{H}_M - \mu \hat{M}) } \right ),
\label{eq:3}
\end{equation}
where $\beta$ and $\mu$ indicate the inverse temperature and chemical potential respectively. 
For a set of parameters $\lambda$, $\mu$ and $\beta$, a corresponding state in thermodynamic equilibrium is 
well defined as long as $\Xi$ converges to a positive real number. One way of ensuring convergence is by requiring 
that the terms having large $M$ in (\ref{eq:3}) go to zero fast enough as $M$ goes to infinity.
A convergence analysis can be done using the fact that the system ground-state energy in the thermodynamic limit $E_G$ is known. 
It is in this way found that in order to guarantee the convergence of $\Xi$ 
the interaction parameter must fulfill $\lambda < \lambda_D$, where
\begin{equation}
\lambda_D = -2(1+\mu) \rightarrow \mu = -\left ( 1 + \frac{\lambda_D}{2} \right),
\label{bocachico}
\end{equation}
if $0 < \lambda_D \le 1$, and
\begin{small}
\begin{equation}
\lambda_D = \frac{1}{2}\left( -\mu + \sqrt{\mu^2 -2} \right) \rightarrow \mu = -\left ( \lambda_D + \frac{1}{2 \lambda_D} \right),
\label{mojarra}
\end{equation}
\end{small}
if $\lambda_D > 1$. The range of valid values for which $\lambda_D>0$ corresponds to $\mu<-1$. 
Since $\mu$ must be negative, it can be argued that thermodynamic equilibrium takes place as 
the instability produced by the attractive interaction, which drives the atoms to cluster up uncontrollably, 
is balanced out by the effect of the bath, which rations the number of particles available 
to the system. As the relation between $\lambda_D$ and $\mu$ is well defined in either direction,
the variables can be used interchangeably.
The grand partition function $\Xi$ is  calculated from Eq. (\ref{eq:3}) by numerically
diagonalizing a set of Hamiltonians $\hat{H}_M$ and then using their respective energy spectra
to find the trace for each value of $M$. The maximum $M$ is chosen so that $\Xi$ converges
to a tolerance of $10^{-7}$. The mean value of an observable $\hat{O}_M$ can be computed as 
a weighted average over the ensemble thus

\begin{equation}
\langle \hat{O} \rangle = \frac{\sum_{M=0}^\infty Tr \left ( \hat{O}_M e^{-\beta (\hat{H}_M - \mu \hat{M}) } \right ) }{\Xi}.
\label{guitar}
\end{equation}

\begin{figure}
\includegraphics[width=0.35\textwidth,angle=-90]{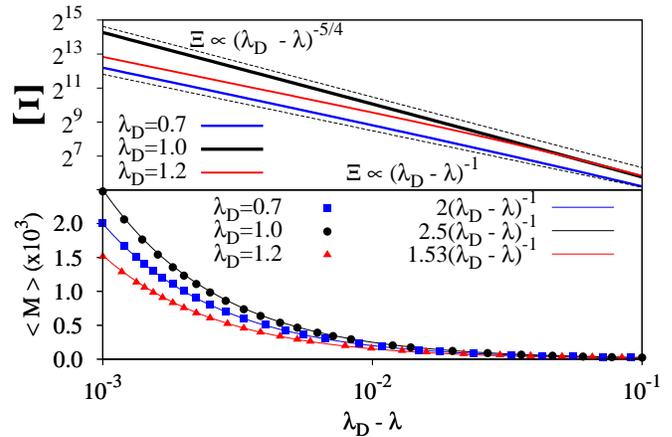}
\caption{In all cases $\beta=1$. Top: A numerical survey shows that the grand canonical 
partition function diverges as $\lambda$ approaches $\lambda_D$ following a power law 
with a characteristic exponent at $\lambda_D=1$. Dashed lines are drawn as a guide to 
the eye. Bottom: The mean number of bosons diverges as the inverse of 
$\lambda_D - \lambda$ multiplied by a coefficient that depends on $\lambda_D$ according 
to (\ref{peru}). The fastest divergence occurs at $\lambda_D=1$.} 
\label{fig1}
\end{figure}

Finding $\Xi$ as well as the mean number of particles and the energy requires only
the eigenenergies. This is computationally faster than finding quantities that
require eigenvectors in addition. Inside the range of parameters for which the sum in Eq. (\ref{eq:3}) converges, 
the numerical simulations show that the grand potential as well as its derivatives are smooth functions of $\lambda$,
even at $\lambda=1$, although they all diverge toward infinity when $\lambda \rightarrow \lambda_D^-$.
This tendency is depicted in figure \ref{fig1} for the case 
of $\Xi$ and the mean number of particles. The lack of discontinuities, and therefore of critical points, 
can be justified by the observation that contributions to the partition function must
decrease with increasing $M$, suppressing in this way the contribution of the limit $M \rightarrow \infty$,
where the original QFT takes place. Since the spectrum of $\hat{H}_M$ is non-degenerate
as long as $M$ be finite, a phase transition in the form of a sudden change in the density of 
states is frustrated. A natural question in this context is whether one can approach the QPT by
standing close to $\beta \rightarrow \infty$ and $\mu = 0$.
This would allow to appreciate how the QPT arises gradually. Interestingly, such an approach
is invalid because $\Xi$ only converges for $\mu<-1$. Hence, the path chosen here is to consider 
the behavior of $\Xi$ when the parameter $\lambda$ is very close to $\lambda_D$ and
the average number of particles is high. 
As shown by figure \ref{fig1}, $\Xi$ diverges following a power law with a peculiar 
feature. The scaling exponent is $-1$ for any valid value of $\lambda_D$, except for 
$\lambda_D=1$, where the exponent becomes $-\frac{5}{4}$. The scaling exponent being
different precisely when $\lambda_D$ equals the critical point of the underlying QFT
suggests that such a point has non-trivial connotations for the finite size
statistics of the problem. Notice that since the grand partition function is a sum of 
terms coming from systems of different size, in this problem the finite size statistics 
has a collective component, involving contributions of independent $\hat{H}_M$. 
\section{Analytic derivation of $\Xi$ near the divergence}
In order to explore the origin of the observed features, let us first consider the
exact grand partition function
\begin{small}
\begin{equation}
\Xi = 1 + \sum_{M=1}^\infty Tr \left( e^{\frac{\beta \lambda}{M} ( \hat{a}_1^\dagger \hat{a}_1 \hat{a}_1^\dagger \hat{a}_1 + \hat{a}_2^\dagger \hat{a}_2 \hat{a}_2^\dagger \hat{a}_2  ) - \beta ( \hat{a}_1^\dagger \hat{a}_2 + \hat{a}_2^\dagger \hat{a}_1 ) + \beta \mu \hat{M} } \right).
\label{eq:9}
\end{equation}
\end{small}
Non-linear terms in the exponential complicate the calculation of the sum
by analytic means. As a workaround, the following linearization scheme is proposed \cite{Niemeijer},    
\begin{widetext}
\begin{equation}
\Xi \approx 1 +\frac{1}{\pi} \sum_{M=1}^\infty \int_{-\infty}^{\infty} dy \int_{-\infty}^{\infty} dx Tr \left( e^{ -(x^2 + y^2) - \sqrt{\frac{4 \beta \lambda}{M}} ( x \hat{a}_1^\dagger \hat{a}_1 + y \hat{a}_2^\dagger \hat{a}_2  ) - \beta ( \hat{a}_1^\dagger \hat{a}_2 + \hat{a}_2^\dagger \hat{a}_1 ) + \beta \mu \hat{M} } \right).
\label{eq:10}
\end{equation}
\end{widetext}
The connection between (\ref{eq:9}) and (\ref{eq:10}) follows from the identity
$\int_{-\infty}^{\infty} e^{-p^2 x^2 - q x}dx = \frac{\sqrt{\pi}}{|p|} e^{\frac{q^2}{4 p^2}}$. 
This step can be seen as a mean field approach and is justified by the
observation that finite size properties close to the thermodynamic
limit are well captured by a mean field treatment \cite{Botet}. 
Since this approximation gets better for large $M$, it is suitable to make 
estimations of $\Xi$ close to the divergence point. The trace and the
sum in Eq. (\ref{eq:10}) can be worked out analytically because the
exponential's argument is a linear operator. The sum converges when 
$\lambda<\lambda_D$, in which case the result can be written as
\begin{equation}
\Xi(\lambda,\beta,\mu) = 1 + \Xi_1(\lambda,\beta,\mu) + \Xi_2(\lambda,\beta,\mu),
\end{equation}
where
\begin{widetext}
\begin{equation}
\Xi_{1 \atop 2} \approx \frac{1}{8 \pi } \int_{-\infty}^{\infty} dw \int_{-\infty}^{\infty} dv
 \frac{ \text{csch} \left ( \frac{1}{2} \left ( \frac{v^2 + w^2}{2} - v \sqrt{\beta \lambda} - \mu \beta  \mp \sqrt{\beta (\lambda w^2 + \beta)}  \right) \right )^2 } { 1 - e^{\mp 2 \sqrt{\beta (\lambda w^2 + \beta)}}}. 
\end{equation}
\end{widetext}
As $\lambda \rightarrow \lambda_D^-$, only $\Xi_1$ diverges. Close to $\lambda_D$ the {\it csch} can be expanded
and the resulting expression can be integrated, thus yielding
\begin{equation}
\Xi_1 \approx  \int_{-\infty}^{\infty} dw \frac{\left( w^2 - 2 \sqrt{\beta(\lambda w^2 + \beta) } - 2  \beta ( \mu + \frac{\lambda}{2} ) \right )^{-3/2}} { 1 - e^{-2\sqrt{\beta(\lambda w^2 + \beta)}} }. \nonumber
\end{equation}
To approximate this integral, the exponential term is kept constant and the
rest of the denominator is expanded in power series in $w$ around the position of its minima. 
For $\lambda \le 1$, there is only one minimum at $w = 0$ and the first three non vanishing terms of the 
expansion are kept. For $\lambda>1$, there are two equal minima located at 
$w=\pm \sqrt{\beta (\lambda - 1/\lambda)}$ and only the first two non vanishing terms of the expansion are required. 
The resulting expression can be solved and the scaling behavior close to 
$\lambda_D$ can be worked out from the solution. Assuming that the original grand partition function
displays the same scaling properties than the approximated expression, it is possible
to write $\Xi$ as follows. If $\lambda_D \le 1$ then
\begin{equation}
\Xi = 1 + \frac{\xi_{\le}(\lambda,\lambda_D,\beta)}{(\lambda_D - \lambda) (1 - \lambda)^{\frac{1}{4}}} + \Xi_2(\lambda,\lambda_D,\beta).
\label{pera}
\end{equation}
Otherwise, when $\lambda_D > 1$
\begin{equation}
\Xi = 1 + \frac{\xi_>(\lambda,\lambda_D,\beta)}{(\lambda_D - \lambda)} + \Xi_2(\lambda,\lambda_D,\beta).
\label{mango}
\end{equation}
In such a way that the limits of $\xi_{\le}$, $\xi_>$ and $\Xi_2$ as $\lambda \rightarrow \lambda_D^-$ 
keeping $\beta$ constant are all finite. The procedure also provides the following estimates
\begin{equation}
\xi_{\le} \approx \frac{\pi( \beta (1 - e^{-2\beta}))^{-1}}{\sqrt{ \sqrt{\frac{3}{2} \left( 1-\lambda + \frac{\lambda^2 (\lambda_D - \lambda)}{1-\lambda }   \right)} + \frac{3}{2}\sqrt{1-\lambda} }},
\label{uva}
\end{equation}
and
\begin{equation}
\xi_> \approx \frac{\lambda^2 \left( \frac{\pi}{2} + \arctan \left ( \sqrt{ \frac{3(\lambda^2 - 1)^2 }{4 \lambda^2 (\lambda_D-\lambda)\left( \lambda - \frac{1}{2 \lambda_D} \right)  }     } \right ) \right ) }{ \beta (1 - e^{-2 \beta \lambda}) \sqrt{\frac{3}{2}(\lambda^2 -1)}  \left( \lambda - \frac{1}{2 \lambda_D} \right)  }.
\label{sandia}
\end{equation}
Owing to the nature of the approximations involved in the derivation, it is expected that these estimations
work better for values of $\lambda$ close to $\lambda_D$ and as such they could reproduce the correct scaling
of thermodynamic variables in such a limit.
\section{repercussions on thermodynamic observables}
The grand partition function determines in great measure the system's statistics
and a number of mean values can be obtained as derivatives of the grand potential. 
Simple relations between important observables can be formulated by focusing on
their behavior near the divergence, which is dominated by a few terms.
Using (\ref{pera}), (\ref{mango}), (\ref{uva}) and (\ref{sandia}) it can be shown
that close to $\lambda_D$ the mean number of particles, which is found from 
$\langle \hat M  \rangle = \frac{1}{\beta} \frac{\partial \log \Xi }{\partial \mu}$,
scales as
\begin{gather}
\begin{array}{cc}
\langle \hat{M}  \rangle \underset {\lambda \rightarrow \lambda_D^-}{\rightarrow}
\begin{cases}
 2 \frac{1}{\beta} (\lambda_D-\lambda)^{-1}                                 & \text{ if } \lambda_D<1. \\
 \frac{5}{2} \frac{1}{\beta} (\lambda_D-\lambda)^{-1}                     &  \text{ if } \lambda_D=1. \\
 \left( 1 - \frac{1}{2 \lambda_D^2} \right)^{-1} \frac{1}{\beta} (\lambda_D-\lambda)^{-1}  & \text{ if } \lambda_D>1.    
\end{cases}
\end{array}
\label{peru}
\end{gather}
This result can be checked against a numerical evaluation of Eq. (\ref{guitar})
taking $\hat{O}_M = \hat{M}$. The bottom panel of figure \ref{fig1} shows that
the two approaches coincide well for values of $\lambda_D-\lambda$ below $0.1$
in three representative cases. The fact that the divergence coefficient changes
discontinuously from $2$ to $2.5$ at $\lambda_D=1$ underscores a trait that could
prove useful as a way of verification of the critical point of the underlying QPT
at any temperature. Analogously, the coefficient reaches its maximum value at this 
point, allowing for an enhanced transfer of particles from the environment to the system.
\begin{figure}[H]
\includegraphics[width=0.35\textwidth,angle=-90]{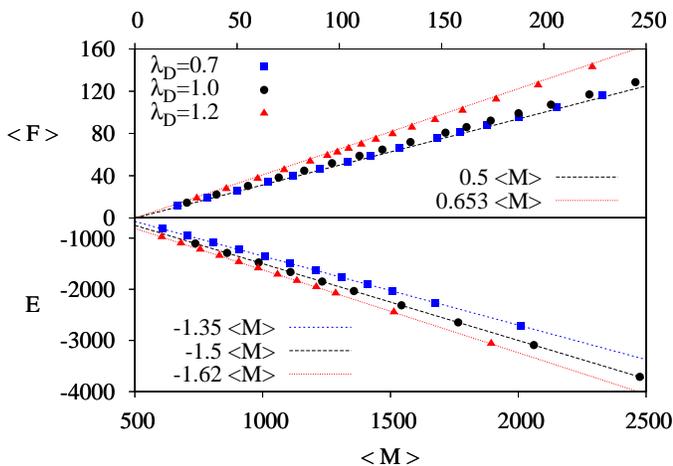}
\caption{In all cases $\beta=1$. Top: Expectation value of the interaction, as defined
in Eq. (\ref{aute}), against the mean number of particles. In this particular 
case the numerical calculation requires the complete eigensystems of various $\hat{H}_M$ and the simulation
does not reach as big $\langle \hat {M} \rangle$ as in the graph below. The fittings correspond to 
the expressions given in (\ref{brasil}), obtained analytically. Bottom: For values of 
$\langle \hat{M} \rangle$ above some hundreds of particles, the relation between 
$\langle \hat{M} \rangle$ and the energy is linear and fits well to Eq. (\ref{chavo}).} 
\label{fig2}
\end{figure}
It is also relevant to consider the behavior displayed by the
interaction term
\begin{equation}
\hat{F}_M = \frac{\hat{a}_1^{\dagger}\hat{a}_1 \hat{a}_1^{\dagger}\hat{a}_1 + \hat{a}_2^{\dagger}\hat{a}_2 \hat{a}_2^{\dagger}\hat{a}_2}{M}.
\label{aute}
\end{equation}
The mean value associated to this quantity could be measured, for example, by counting the number of particles
in each mode, $n_1$ and $n_2$, and then averaging $(n_1^2 + n_2^2)/(n_1+n_2)$ over
measurements obtained from identically prepared setups having the same $\lambda$, $\lambda_D(\mu)$ and $\beta$. 
The interaction mean-value can be obtained from the analytical estimation of $\Xi$ as 
$\langle \hat F  \rangle = \frac{1}{\beta} \frac{\partial \log \Xi }{\partial \lambda}$.
Near the divergence the resulting expression can be written as
\begin{gather}
\begin{array}{cc}
\langle \hat{F} \rangle \underset {\lambda \rightarrow \lambda_D^-}{\rightarrow}
\begin{cases}
 \frac{\langle \hat{M} \rangle}{2}                                   & \text{ if } \lambda_D \le 1. \\
\left( 1 - \frac{1}{2 \lambda_D^2} \right)  \langle \hat{M} \rangle  & \text{ if } \lambda_D>1.    
\end{cases}
\end{array}
\label{brasil}
\end{gather}
Where in every case $\langle \hat{M} \rangle$ is given by (\ref{peru}).
A comparison with numerical results obtained by direct application of
Eq. (\ref{guitar}) taking $\hat{O}_M = \hat{F}_M$
can be seen in the top panel of figure \ref{fig2}. Since this computation
requires eigenvectors, it is not as easy to reach bigger values of $\langle \hat{M} \rangle$ 
for which a better agreement between analytics and numerics is expected. 
The proportionality coefficient between $\langle \hat{F} \rangle$ and 
$\langle \hat{M}  \rangle$ is a continuous, but not smooth, function of $\lambda_D$
at $\lambda_D=1$. The system's energy can also be obtained from $\Xi$ as
$E = \mu \langle \hat M \rangle - \partial \log \Xi / \partial \beta$. In
this case differentiation with respect to $\beta$ does not contribute divergence
terms and the energy becomes simply       
\begin{equation}
E \underset {\lambda \rightarrow \lambda_D^-}{\rightarrow} \mu \langle \hat M \rangle.
\label{chavo}
\end{equation}
It can be checked using (\ref{bocachico}) and (\ref{mojarra}) that the 
proportionality coefficient ($\mu$) and its derivative are continuous functions of $\lambda_D$.
A comparison with a direct application of Eq. (\ref{guitar}), this time making $\hat{O}_M = \hat{H}_M$, can be seen in 
the lower panel of figure \ref{fig2}. It becomes in this way apparent that the functionality 
of $\Xi$ with respect to the problem's parameters has been correctly captured by Eqs. (\ref{pera}), 
(\ref{mango}), (\ref{uva}) and (\ref{sandia}) in the region close to the divergence
and that there exist features that would allow to identify the point $\lambda_D=1$
in a practical scenario. 
\section{Discussion}
Let us now go back the fact that the QPT studied here, which is formulated in 
connection to canonical-ensemble theory in section \ref{intro}, cannot be seen as an emergent
phenomenon in the thermodynamics resulting from the grand canonical
formalism in section \ref{gca}. This might appear a little odd since it is known from numerics 
that Hamiltonians $\hat{H}_M$ with large $M$ display results that asymptotically approach
those in the thermodynamic limit. Although no attempt will be made here to provide a simple
justification for this behavior, it can be shown that in this case the canonical formalism and
the grand canonical formalism do not need to coincide in the thermodynamic limit. In many
cases an overlap takes place between the microcanonical- and the canonical ensemble because
in the latter the energy scales as the number of particles while fluctuations scale as its 
square root, so that the energy becomes more defined for big occupations. To check for a 
potential equivalence in the current case, it is necessary to look at the fluctuations in 
the number of particles, which can be obtained from
\begin{equation}
\sigma_M^2 = \langle \hat{M}^2 \rangle - \langle \hat{M} \rangle^2 = \frac{1}{\beta^2} \frac{\partial^2 \log \Xi}{\partial \mu^2}.
\end{equation}
Proceeding as in the previous section, it follows
\begin{gather}
\begin{array}{cc}
\sigma_M \underset {\lambda \rightarrow \lambda_D^-}{\rightarrow}
\begin{cases}
\langle \hat{M} \rangle                                 & \text{ if } \lambda_D \ne 1. \\
\sqrt{\frac{4}{5}} \langle \hat{M} \rangle  & \text{ if } \lambda_D=1.    
\end{cases}
\end{array}
\label{china}
\end{gather}
Therefore, fluctuations are of the same order than $\langle \hat{M} \rangle$ and 
the number of particles remains a statistical parameter, in contrast to the
case in the canonical formalism, where the number of particles is fixed.
\section{Conclusion}
A simple many-particle model has been introduced and shown to display a QPT 
in accordance to canonical ensemble theory. A detailed study of the same 
structure with reference to the macrocanonical ensemble shows no sign of
a phase transition nor a way to approach the canonical ensemble conditions.
The grand canonical partition function displays a divergence with relation 
to the interaction parameter which follows a
power law with an exponent that takes a characteristic value in a case
clearly related to the critical point of the underlying QPT. The scaling
properties of the grand partition function, and from it the power law exponents, 
were determined analytically using a mean field formulation. It is shown that
the observed features affect the thermodynamic observables in a way that makes it
possible to identify the critical point of the underlying QFT from relations
involving thermodynamic observables and Hamiltonian parameters. It remains 
to be seen whether similar signatures are displayed by models undergoing 
phase transitions at finite temperature, so that a critical point is approached 
by shifting the temperature rather than a Hamiltonian parameter. Due to their 
similarity with Hamiltonian (\ref{eq:1}), the LMG model and its correlative, 
the Dicke model \cite{Reslen1,Vidal}, seem suitable scenarios to explore the 
thermodynamic response connected to the grand canonical formalism.
Evidence seems to suggest that the precursors of a QPT in the thermodynamic
limit stay active, albeit to a lesser extent, in the collective response of 
the finite-size profile of a system.  

The authors thank the support of Vicerrector\'ia de Investigaciones,
Extensi\'on y Proyecci\'on Social from Universidad del Atl\'antico.

\begin{thebibliography}{}
\bibitem{comment1} Here only the ground state degeneration is shown. Numerical simulations show that this also happens to the rest of the spectrum.
\bibitem{comment2} At zero temperature the free energy is just $f=E_G$.
\end{thebibliography}
\end{document}